\documentclass{pasj00_arxiv}
\newcommand{\bm}[1]{\mbox{\boldmath $#1$}}

\begin{document}
\SetRunningHead{M. Nomura et al.}{Line-Driven Disk Winds for Ultra Fast Outflows}
\Accepted{for publication in PASJ}

\title{Radiation Hydrodynamic Simulations of Line-Driven Disk Winds for Ultra Fast Outflows}



%
\author{%
  Mariko \textsc{Nomura}\altaffilmark{1}
  Ken \textsc{Ohsuga}\altaffilmark{2,3}
  Hiroyuki R. \textsc{Takahashi}\altaffilmark{2}
  Keiichi \textsc{Wada}\altaffilmark{4}
  and
  Tessei \textsc{Yoshida}\altaffilmark{2}}
\altaffiltext{1}{University of Tsukuba, Center for Computational Sciences, Tennodai, Tsukuba, Ibaraki 305-8577}
\altaffiltext{2}{National Astronomical Observatory of Japan, Osawa, Mitaka, Tokyo 181-8588}
\altaffiltext{3}{School of Physical Sciences, Graduate University of
Advanced Study (SOKENDAI), Shonan Village, Hayama, Kanagawa 240-0193}
\altaffiltext{4}{Graduate School of Science and Engineering, Kagoshima University, Kagoshima 890-0065}

\email{nomura@ccs.tsukuba.ac.jp}


\maketitle

\begin{abstract}
Using two-dimensional radiation hydrodynamic simulations, 
we investigate origin of the ultra fast outflows (UFOs) 
that are often observed in luminous active galactic nuclei (AGNs). 
We found that the radiation force due to the spectral lines generates 
strong winds (line-driven disk winds) 
that are launched from the inner region of accretion disks ($\sim 30$ Schwarzschild radii).  
A wide range of black hole masses ($M_{\rm BH}$) 
and Eddington ratios ($\varepsilon$) was investigated to study conditions 
for causing the line-driven winds. 
For $M_{\rm BH} = 10^6$--$10^9 M_\odot$ and $\varepsilon = 0.1$--$0.7$, 
funnel-shaped 
disk winds appear, 
in which dense matter is accelerated outward with an opening angle of 
70--$80^\circ$ and with 10\% of the light speed.  
If we observe the wind along its direction, the velocity, the column density, 
and the ionization state are consistent with those of the observed UFOs. 
As long as the obscuration by the torus does not affect the observations of X-ray bands, 
the UFOs could be statistically observed in about 13--28\% of 
the luminous AGNs, 
which is not inconsistent with the observed ratio ($\sim 40\%$).
We also found that the results are insensitive to 
the X-ray luminosity and the density of the disk surface. 
Thus, we can conclude that the UFOs could exist in any luminous AGNs, 
such as narrow-line Seyfert 1s (NLS1s) and quasars with $\varepsilon > 0.1$, 
in which fast line-driven winds are associated.

\end{abstract}

\section{Introduction}
Outflows are frequently observed in active galactic nuclei (AGNs) and 
their origin has been discussed.
The blueshifted absorption lines in the spectra suggest the existence of the outflows.
Typical absorption lines are broad absorption lines (BALs), 
of which blueshifted velocities are $\sim 10,000\,{\rm km\, s^{-1}}$ and the line widths exceed $2,000\,{\rm km\, s^{-1}}$ \citep{We91,Hamann93,Gibson09,Allen11}. 
BALs are observed in optical or ultraviolet (UV) spectra and detected in 10--20\% of quasars.
The blueshifted absorption lines are observed also in X-ray band (e.g., \cite{Pounds03,Tombesi10}). 
Such absorption features are thought to be made by FeX\hspace{-.1em}X\hspace{-.1em}V and/or FeX\hspace{-.1em}X\hspace{-.1em}V\hspace{-.1em}I
in the outflow.
The velocity and the column density of the outflow are
roughly evaluated as $0.1$--$0.3c$ and $10^{22}$--$10^{24}\,{\rm cm^{-2}}$,
where $c$ is the speed of light.
These high-velocity outflows are called ultra fast outflows (UFOs).
Though the number of the samples is still small, UFOs are detected in about a half of Seyfert galaxies. 
Since the kinetic luminosity of the UFOs is evaluated to be comparable to or larger than that of jets
\citep{Tombesi12},
the UFOs may contribute to the AGN feedback.
In addition, the mass outflow rate of UFOs 
is at least 5--10\% of the mass accretion rate onto the black holes.
Thus, UFOs may have an impact on the growth of the super massive black holes.
However, the origin of UFOs is still unknown. 

The dichotomy between AGNs with the absorption lines and those without the absorption lines 
is the problem still in discussion. 
\citet{We91} suggested that the difference between them is caused by the viewing angle of the observer, 
since the properties except for the absorption lines are quite similar between the BAL and non-BAL quasars.
\citet{Elvis00} proposed the funnel-shaped disk wind model.
In this model, if the wind interrupts the observer's line of sight
(the observer's viewing angle coincides with the opening angle of the wind),
the blueshifted absorption lines appear in the spectra. 
In contrast, 
if there is no wind between the nuclei and the observer, 
the absorption features are not detected.
On the other hand, 
\citet{Zu13} have proposed another idea that the outflows intermittently appear.
In this case, the presence or absence of the absorption features
changes with time independently of the viewing angle.

Until now, many theoretical models have been proposed to explain the origin of the outflow.
One of the most plausible models is the line-driven disk winds, 
in which radiation force due to spectral lines
(the absorbing UV radiation through the bound-bound transition of metals)
accelerates the matter and launches the outflows from the disk surface.
This mechanism is effective only for 
the matter in low- or moderate-ionization state,
since the bound-bound transition cannot occur 
if the metals are fully ionized. 
Thus, the line-driven wind model is capable of explaining
the acceleration and the ionization state of the wind,
which are required to reproduce the blueshifted absorption lines,
at the same time.
The discovery of the line locking in some quasars \citep{F87} indicates 
that the line force can play an important role for acceleration of the matter.
\citet{SK90} (hereafter SK90) showed that the line-driven mechanism becomes more effective for the matter 
in the low-ionization state and the radiation force including the line force 
can be $10$--$1000$ times larger than the radiation force due to the electron scattering.
It implies that the line-driven wind can be launched from the sub-Eddington disks.
The magnetically driven winds
can also explain the acceleration of the outflows
\citep{B82,K94,Ever07}.
However, the extra mechanism is required to 
realize the optimum ionization state of the metals.

The radiation hydrodynamic simulations of the line-driven winds
launched from the disks around supermassive black holes 
have been performed by \citet{Proga00} and \citet{Proga04} 
(hereafter PSK00 and PK04).
They improved the numerical method for the line-driven winds
around white dwarfs 
(\authorcite{Proga98} \yearcite{Proga98}, \yearcite{Proga99}),
and applied to the outflows in AGNs.
%
%
PSK00 and PK04 
reproduced
the funnel-shaped disk wind.
The wind comes from a distance from the black hole of $\sim$ several
$100R_{\rm S}$ and goes away in the direction of $\theta\sim 70^{\circ}$
for the typical parameters, $M_{\rm BH}=10^8 M_{\odot}$ and
$\varepsilon=0.5$, where $R_{\rm S}$ is the Schwarzschild radius,
$\theta$ is the polar angle measured from the rotation axis of the disk,
$M_{\rm BH}$ is the black hole mass, and $\varepsilon$ is the Eddington
ratio of the UV radiation.
Based on the results of PK04,
\citet{Schurch09} and \citet{Sim10} performed the spectral synthesis 
and compared them to the X-ray observations.

Although PSK00 and PK04 mentioned that
the winds are produced for the range of $M_{\rm BH}>10^7 M_{\odot}$ 
and the winds do not appear for $\varepsilon=0.1$,
their simulations focus on the case of 
$M_{\rm BH}=10^8 M_{\odot}$ and $\varepsilon=0.5$.
The line-driven disk wind 
in the wide parameter space of the black hole mass and the Eddington
ratio was investigated by \citet{Risaliti10}. 
They used a non-hydrodynamic method, 
in which the trajectories of the fluid elements 
are calculated by solving the equation of motion including the line force,
and 
found
the steady structure of the line-driven winds.
Using the similar non-hydrodynamic method, 
\citet{Nomura13} also 
investigated the wind properties, such as the ionization state, 
the outflow velocity, and the column density 
and compared them to those inferred from the X-ray observations of BALs. 
As a result, 
it has been concluded 
that the BALs are detected in 10\% of AGNs
in the case of $M_{\rm BH}\sim 10^7$--$10^8 M_{\odot}$ 
and $\varepsilon\sim 0.3$--$0.9$.

Thus, previous studies focused either on only a typical case using hydrodynamic simulations, or on kinematics of the wind in a wide parameter space without considering hydrodynamic effects.
However, the radiation hydrodynamic simulations in the wide parameter space 
have not been performed yet. 
In the previous studies,
the line-driven wind model 
has never been applied to UFOs,
although UFOs are detected 
in a wide variety of the black hole mass and the Eddington ratio 
\citep{Tombesi11}.
Since the UFOs are important with respect to 
the evolution of super massive black holes and 
the feedback to the host 
galaxies, 
we here try to understand origin of the UFOs in a context of the line-driven wind model.
In this paper, 
we investigate the line-driven winds for various black hole masses
and the UV luminosities
by performing the two-dimensional radiation hydrodynamic simulations.
We also investigate the cases that the X-ray photoionization is more effective
and the density at the disk surface is relatively large or small.
In addition, 
we calculate the ionization parameter, 
the outflow velocity, and the column density 
along the line of sight. 
Then, by comparing them to the observational features of the UFOs,
we examine whether or not the line-driven wind 
may explain the phenomena.
We explain our calculation method in section \ref{method}. 
We present results of the simulations and comparison with the
observational features in  section \ref{results}. 
We devote  section \ref{discussions} and  section \ref{conclusions} 
to discussion and conclusions.


\section{Method}
\label{method}
\subsection{Basic equations and set up}
\label{method1}
The method of the calculation is almost similar to that of PSK00 and PK04.
We use spherical polar coordinates $(r, \theta, \varphi)$, where $r$ is the distance from the origin of the coordinate, $\theta$ is the polar angle, and $\varphi$ is the azimuthal angle. 
We assume axial symmetry with respect to the rotation axis of the accretion disk.
The basic equations of the hydrodynamics are the equation of continuity,
\begin{equation}
\frac{\partial \rho}{\partial t}+\nabla\cdot(\rho \mbox{\boldmath $v$})
=0,
\label{eoc}
\end{equation}
the equations of motion,
\begin{equation}
\frac{\partial (\rho v_r)}{\partial t}+\nabla\cdot(\rho v_r \mbox{\boldmath $v$})
=-\frac{\partial p}{\partial r}+\rho\Bigg[\frac{v_\theta^2}{r}+\frac{v_\varphi^2}{r}+g_r+f_{{\rm rad},\,r}\Bigg],
\label{eom1}
\end{equation}
\begin{equation}
\frac{\partial (\rho v_\theta)}{\partial t}+\nabla\cdot(\rho v_\theta \mbox{\boldmath $v$})
=-\frac{1}{r}\frac{\partial p}{\partial \theta}+\rho\Bigg[-\frac{v_r v_\theta}{r}+\frac{v_\varphi^2}{r}\cot \theta+g_\theta+f_{{\rm rad},\,\theta}\Bigg],
\label{eom2}
\end{equation}
\begin{equation}
\frac{\partial (\rho v_\varphi)}{\partial t}+\nabla\cdot(\rho v_\varphi \mbox{\boldmath $v$})
=-\rho\Bigg[\frac{v_\varphi v_r}{r}+\frac{v_\varphi v_\theta}{r}\cot \theta\Bigg],
\label{eom3}
\end{equation}
and the energy equation,
\begin{equation}
\frac{\partial}{\partial t}\Bigg[  \rho \Bigg(\frac{1}{2}v^2+e\Bigg) \Bigg]
+\nabla \cdot \Bigg[  \rho\mbox{\boldmath $v$} \Bigg(\frac{1}{2}v^2+e+\frac{p}{\rho}\Bigg) \Bigg]
=\rho\mbox{\boldmath $v$}\cdot\mbox{\boldmath $g$}+\rho\mathcal L,
\label{eoe}
\end{equation}
where $\rho$ is the mass density, 
\mbox{\boldmath $v$}$=(v_r,\,v_\theta,\,v_\varphi)$ is the velocity, $p$ is the gas pressure,
$e$ is the internal energy per unit mass, 
and \mbox{\boldmath $g$}$=(g_r,\,g_\theta )$ is the gravitational acceleration of the black hole.  
Since the origin 
of the coordinate
($r=0$) in the present simulations
is slightly distant from the center of the black hole,
the $\theta$-component of the gravitational force is not null.
We will explain about the computational domain at the end of this subsection.
We assume an adiabatic equation of state $p/\rho=(\gamma -1)e$ 
with $\gamma=5/3$.

In the last terms of the equations (\ref{eom1}) and (\ref{eom2}), 
\mbox{\boldmath $f$}$_{\rm rad}=(f_{{\rm rad},\,r},\,f_{{\rm rad},\,\theta})$ 
is the radiation force per unit mass including the line force,
\begin{equation}
{ \mbox{\boldmath $f$}_{\rm rad}}=\frac{\sigma_{\rm e} \mbox{\boldmath $F$}_{\rm D}}{c}+\frac{\sigma_{\rm e} \mbox{\boldmath $F$}_{\rm D}}{c}M,
 \label{radforce}
\end{equation}
where 
$\sigma_{\rm e}$ is the mass-scattering coefficient for free electrons,  
${\bm F}_{\rm D}$ is the flux from the accretion disk, 
and $M$ is the force multiplier 
(see section \ref{section:forceM}).
The first term is the radiation force due to the electron scattering and the second term is the line force.
We simply assume that 
all of photon emitted from the hot region of the accretion disk 
is able to contribute to the line force. 
We explain the definition of $F_{\rm D}$ in section \ref{flux}.

%
In the second term on the right-hand side of the equation (\ref{eoe}),
$\mathcal L$ is the net cooling rate (see PSK00),
\begin{equation}
\mathcal L=n^2(G_{\rm Compton}+G_{\rm X}-L_{\rm b,l}),
\end{equation}
where, $n$ is the number density, $G_{\rm{Compton}}$ is Compton heating/cooling rate,
\begin{eqnarray}
G_{\rm{Compton}}=8.9\times10^{-36}\xi(T_{\rm X}-4T),
\end{eqnarray}
$G_{\rm X}$ is the rate of X-ray photoionization heating and recombination cooling,
\begin{eqnarray}
G_{\rm X}=1.5\times10^{-21}\xi^{1/4}T^{-1/2}(1-T/T_{\rm X}),
\end{eqnarray}
$L_{\rm{b,l}}$ is the bremsstrahlung and line cooling,
\begin{eqnarray}
\nonumber L_{\rm{b,l}}=3.3\times10^{-27}T^{1/2}+1.7\times10^{-18}\xi^{-1}T^{-1/2}\\
\times \exp(-1.3\times10^5/T)+10^{-24}.
\end{eqnarray}
%
In above three equations, $T$ is the temperature obtained by
$T=\mu m_{\rm p} p/k_{\rm B}\rho$, 
where $\mu$ (=0.5) is the mean molecular weight,
$m_{\rm p}$ is the proton mass,
and $k_{\rm B}$ is the Boltzmann constant.
The ionization parameter is denoted by $\xi$ 
defined as 
\begin{equation} 
 \xi=\frac{4\pi F_{\rm X}}{n},
  \label{xi}
\end{equation}
where $F_{\rm X}$ is the X-ray flux 
(see section \ref{flux} for detailed explanation).
%
We assume the temperature of the X-ray radiation to be
$T_{\rm X} =10^8\,\rm{K}$,
following PSK00.

As with PK04,
our computational domain occupies the radial range of $r_{\rm i} = 30R_{\rm S}\leq r\leq r_{\rm o}=1500R_{\rm S}$ and the angular range of $0^{^\circ } \leq \theta \leq 90^{\circ}$.
The plane of $\theta=90^{\circ}$ does not correspond to the equatorial plane of
the accretion disk, but is located at the surface of the accretion disk.
Specifically, this plane is placed in parallel with the equatorial plane
at a distance of $z_0$.
Here, 
we set $z_0=4.0 \varepsilon R_{\rm S}$,
which roughly corresponds to the scale height of the standard disk
\citep{SS73}.
Throughout the present study,
we employ the standard disk model,
which is a source of the radiation and 
supplies the mass to the disk wind.
Since $z_0$ is estimated 
based on the hydrostatic equilibrium in the vertical direction,
it is proportional to the Eddington ratio \citep{Kato_textbook}.
We divide this computational domain into 100 grids for the radial range 
and 160 grids for the angular range.
To get the good spatial resolution 
at the inner region and near the disk surface,
we use the fixed zone size ratios, 
$\Delta r_{j+1}/\Delta r_{j}=1.05$ 
and $\Delta\theta_{k}/\Delta\theta_{k+1}=1.066$.
The origin of our coordinate is 
located at $4.0 \varepsilon R_{\rm S}$ above the center of the black hole,
so that 
the $\theta$-component of the gravitational force is not null
(see the equation [\ref{eom2}]).
Here, we note that the resulting wind structure does not change so much 
if we prepare 150 grids for the radial range, 
in which case the resolution is 4--10 times better than that for 100 grids 
at $r\sim 30$--$40 R_{\rm S}$, which corresponds the launching region of the wind.

\subsection{Boundary conditions}
\label{BC}
We set boundary conditions as follows. 
We apply the axially symmetric boundary 
at the rotational axis of the accretion disk, $\theta=0^{\circ}$
($\rho$, $p$, and $v_r$ are symmetric,
while $v_\theta$ and $v_\varphi$ are antisymmetric). 
The outflow boundary conditions are employed 
at the inner and outer radial boundaries,
at $r=r_{\rm i}$ and $r_{\rm o}$.
All quantities are basically symmetric, 
but the matter can freely go out but not to come in 
the computational domain.
%
At the boundary of $\theta=90^{\circ}$, 
we apply the reflecting boundary
($\rho$, $p$, $v_r$, and $v_\varphi$ are symmetric,
but $v_\theta$ is antisymmetric). 

At the boundary of $\theta=90^{\circ}$, 
the radial velocity and the rotational velocity
are fixed to be null and the Keplerian velocity.
Also, the density is kept constant of $\rho=\rho_0$,
and the temperature is fixed on the effective temperature 
of the standard disk surface. 
The density, $\rho_0$, corresponds to the density 
at the photosphere of the accretion disk.
Basically, we employ $\rho_0=10^{-9}\,{\rm g\,cm^{-3}}$.
We also investigate other cases 
of $\rho_0=10^{-11}$ and $10^{-7}\,{\rm g\,cm^{-3}}$
for comparison (see Appendix \ref{App2}).
In reality, the density at the disk surface depends on the black hole mass, the Eddington ratio and the disk radius. 
However, in this paper, 
we set $\rho_0$ to be constant with the disk radius as a free parameter.

\subsection{Initial conditions}
\label{IC}
We set the initial conditions based on \citet{Proga98} and PSK00.
We assume hydrostatic equilibrium in the vertical direction, then the initial density distribution is 
\begin{equation}
\rho(r, \theta)=\rho_0 \exp \Biggl( -\frac{GM_{\rm BH}}{2c_{\rm s}^2 r\tan^2 \theta} \Biggr),
\end{equation}
where $c_{\rm s}$ is the sound velocity at the disk surface.
The initial temperature at a given point, $(r,\theta)$,
is set to be $T(r, \theta)=T_{\rm eff}(r\sin\theta)$,
where $T_{\rm eff}$ is the effective temperature of the disk
(see the equation [\ref{disktemp}]),
i.e.,
there is no temperature gradient in the vertical direction.
%
%
%
We set the initial velocity to be $v_{r}=v_{\theta}=0$. 
The rotational velocity, $v_{\varphi}$, is given 
to meet the equilibrium between the gravity and the centrifugal force
as $v_{\varphi}=\left( GM_{\rm BH}/r \right)^{1/2} \sin\theta$. 

%

\subsection{Line force}
\label{hydro-lineforce}
\subsubsection{Radiative fluxes emitted from central X-ray source and accretion disk}
\label{flux}

In our simulations, 
we consider two radiation sources,
a point source located at the center of the coordinate 
and hot region of the accretion disk 
located at the equatorial plane.
The radiation from the central point source is assumed 
to contribute the ionization of the metals and called X-ray in this paper. 
We assume that all photons emitted from the high-temperature region of the disk
contribute to the line force and 
note that the ionization due to the high-energy photons 
(photons of which energy is above $13.6\,{\rm eV}$) 
emitted from the disk is not taken into account. 
To estimate the ionization parameter defined by the equation (\ref{xi}), 
we use the X-ray flux that we calculate here.
The luminosities of X-ray source and the accretion disk are described as $L_{\rm X}=f_{\rm X}\varepsilon L_{\rm Edd}$ and $L_{\rm D}=\varepsilon L_{\rm Edd}$,
where $\varepsilon$ is the Eddington ratio of the disk luminosity and 
$f_{\rm X}$ is the ratio of the X-ray luminosity to the disk luminosity.
Basically, we set $f_{\rm X}=0.1$, 
but we employ $f_{\rm X}=1$ in section \ref{XUV}
in order to investigate the wind 
when the X-ray photoionization is so effective.
%
The disk radiation comes from
the geometrically thin and optically thick disk 
(standard disk) whose 
effective temperature distribution is given by
\begin{equation}
 T_{\rm eff}(r)=T_{\rm in} \left( \frac{r}{r_{\rm in}} \right)^{-3/4},
\label{disktemp}
\end{equation}
where $r_{\rm in}(=3R_{\rm S})$ is the disk inner radius and $T_{\rm in}$ is the effective temperature at $r=r_{\rm in}$.
We set $T_{\rm in}$ to meet the condition of
\begin{equation}
L_{\rm D}=\varepsilon L_{\rm Edd}=\int_{r_{\rm in}}^{r_{\rm out}}
2\pi r \sigma T_{\rm eff}^4 dr,
\label{T_in}
\end{equation}
where 
$\sigma$ is the Stefan-Boltzmann coefficient, and 
$r_{\rm out}$ is the outer radius of 
the inner hot region,
where we suppose $T_{\rm eff}(r_{\rm out})=3\times10^3 \,\rm K$.
At the region of $r>r_{\rm out}$, 
the temperature is too cold to emit 
photons contributing to the line force
effectively. 
We use the simple method in the same manner as PSK00, 
which supposes that the radiation from the high-temperature ($>3\times 10^3\,{\rm K}$) region 
of the accretion disk contributes to the line force. 
(In PK04, the radiation between $200\,{\rm \AA}$ and $3200\,{\rm \AA}$ is assumed to contribute to the line force.)

Using the X-ray and disk fluxes in the optically thin media,
$F_{\rm X,thin}$ and 
$\mbox{\boldmath $F$}_{\rm D,\,thin}
=(F_{\rm D,thin}^r,F_{\rm D,thin}^\theta)$,
which are computed before starting the simulations,
we evaluate the radial components of the X-ray and disk fluxes as
$F_{\rm X}=F_{\rm X,thin} e^{-\tau_{\rm X}}$ 
and $F_{\rm D}^r=F_{\rm D,thin}^r e^{-\tau_{\rm D}}$.
Here, $\tau_{\rm X}$ and $\tau_{\rm D}$ are the optical depths 
for the X-ray and 
the disk radiation
calculated at each time step.
The dilution of the $\theta$-component of the disk flux 
is supposed to be negligible, 
$F_{\rm D}^{\theta}=F_{\rm D, thin}^{\theta}$.

%
The disk flux
in the optically thin media 
is obtained by
\begin{equation}
  \mbox{\boldmath $F$}_{\rm D,\,thin}
  (r,\theta)=\int \frac{\sigma T_{\rm eff}^4}{\pi}
  \mbox{\boldmath $n$} d\Omega,
  \label{UVflux}
\end{equation}
where \mbox{\boldmath $n$} is the unit vector, 
and $d\Omega$ is the solid angle
from the point $(r, \theta)$ to the 
infinitesimal surface element of the disk.
%
Here, 
the disk radiation is supposed to come from the equatorial plane.
%
The X-ray radiation is spherical symmetry, and the radial component of the X-ray flux in the optically thin media is $F_{\rm X,thin}=L_{\rm X}/4\pi r^2$.

We calculate the optical depths for the X-ray and the radiation from the accretion disk as 
\begin{equation}
\tau_{\rm X}(r,\theta)=\int_{r_{\rm i}}^{r} \sigma_{\rm X}(\xi)\rho(r',\theta) dr',\\
\label{tau_iX}
\end{equation}
and
\begin{equation}
\tau_{\rm D}(r,\theta)=\int_{r_{\rm i}}^{r} \sigma_{\rm e}\rho(r',\theta) dr',
\label{tau_iUV}
\end{equation}
where $\sigma_{\rm X}$ is the mass extinction coefficient for X-ray. 
We set $\sigma_{\rm X}=\sigma_{\rm e}$ for $\xi\ge 10^5$
and $\sigma_{\rm X}=100 \sigma_{\rm e}$ for $\xi< 10^5$
in order to include the effect of the photoelectronic absorption.

\subsubsection{Force multiplier}
\label{section:forceM}
To evaluate the line force, we use the force multiplier 
proposed by SK90, which is modified version of \citet{CAK75}.
The force multiplier is the ratio of the line force to the radiation force due to the electron scattering and written as 
\begin{equation}
  M(t,\xi)=kt^{-0.6}\Biggl[ \frac{(1+t\eta_{\rm{max}})^{0.4}-1}{(t\eta_{\rm{max}})^{0.4}}\Biggr].
  \label{forceM}
\end{equation}
Here, $t$ is the local optical depth parameter,
\begin{equation}
  t=\sigma_{\rm e} \rho v_{\rm{th}}\Bigl| \frac{dv}{ds}\Bigr|^{-1},
   \label{t-xi}
\end{equation}
where $v_{\rm{th}}$ is the thermal speed of the gas,
and $dv/ds$ is the velocity gradient along the light-ray.
In the equation (\ref{forceM}), $k$ and $\eta_{\rm{max}}$ are the functions of the ionization parameter, $\xi$,
and written as
\begin{equation}
  k=0.03+0.385\exp(-1.4\xi^{0.6}),
\end{equation}
and 
\begin{equation}
 \log_{10}\eta_{\rm{max}}=\left\{ 
    \begin{array}{ll}
      6.9\exp(0.16\xi ^{0.4}) & \log\xi \le 0.5 \\
      9.1\exp(-7.96\times 10^{-3}\xi) & \log\xi >0.5 \\
    \end{array} \right.
 .
  \label{k_eta}
\end{equation}

The force multiplier formally depends on the thermal speed through the local optical depth parameter, but in other modified models, the line force is given in a form independent of the thermal speed (e.g., \cite{Gayley95}). 
We set $v_{\rm th}=20\,{\rm km\,s^{-1}}$ in the same way as PSK00, 
which corresponds to the thermal speed of hydrogen gas
whose temperature is $25,000\,{\rm K}$.
The force multiplier depends on the direction of the light-ray
thorough $dv/ds$.
However, in our simulations, 
we approximate the velocity gradient along the light-ray 
by the velocity gradient in the radial direction,
\begin{equation}
 \frac{dv}{ds}\simeq \frac{dv_{r}}{dr}.
  \label{grad_v2}
\end{equation}
By this approximation, the force multiplier becomes independent of the
direction and is determined locally. 
%
By calculating the ionization parameter, $\xi$, and the local optical depth parameter, $t$, we evaluate the force multiplier at each time step.

\subsection{Numerical code}
The numerical procedure is divided into the following steps.
%
(i) 
The optical depths for the 
radiation emitted from the disk
and X-ray are calculated
by solving the equations (\ref{tau_iX}) and (\ref{tau_iUV}).
Using the optical depths,
the radiation fluxes are evaluated (section \ref{flux})
and we calculate the force multiplier (section \ref{section:forceM})
as well as the radiation force including the line force
(the equation [\ref{radforce}]).
(ii) 
The equations (\ref{eoc})--(\ref{eom3}), 
and the energy equation (the equation [\ref{eoe}]) except 
for the net cooling term are calculated.
We treat the radiation force as the external force term explicitly.
Here, hydrodynamic terms of ideal fluid are solved 
with using an approximate Riemann solver, HLL method (Harten et al. 1983).
The numerical code here we use is 
a part of \citet{Takahashi13}.
(iii) 
The temperature (i.e., internal energy of the fluid) is implicitly updated
by considering the net cooling rate
with using the 
bisection method.

At the step (ii), the internal energy
sometimes becomes negative for the case that the velocity is very large.
In that case, 
we evaluate the gas pressure (internal energy) 
from the entropy equation
instead of the energy equation,
\begin{equation}
\frac{\partial s}{\partial t}+\nabla\cdot(s\mbox{\boldmath $v$})
=(\gamma-1)\rho^{2-\gamma}\mathcal L
\label{entropy}
\end{equation}
where $s$ is the entropy density, $s=p/\rho^{\gamma-1}$.
Finally, 
we set the lower limits of the density to be $10^{-22}\,{\rm g\,cm^{-3}}$
and the temperature to be $T_{\rm eff}(r\sin \theta)$.
If the density or the temperature is lower than them,
we replace them with the lower limits.

The time step is determined by using the Courant-Friedrichs-Levi condition.
At each grid, we calculate
\begin{equation}
\Delta t =0.05  \frac{\min(\Delta r,r\Delta\theta)}{\sqrt{(v_r+c_{\rm s})^2+(v_{\theta}+c_{\rm s})^2}}
\end{equation}
where $\Delta r$ and $\Delta\theta$ are the grid sizes in the radial and polar directions.
The minimum value of $\Delta t$ in all grids is used as the time step.

\section{Results}
\label{results}
The resulting wind structures 
for $\varepsilon=0.5$, $M_{\rm BH}=10^8 M_{\odot}$, $f_{\rm X}=0.1$, 
and $\rho_0=10^{-9}\,{\rm g\,cm^{-3}}$ 
(hereafter, we call `fiducial model')
are consistent with the results of PSK00 and PK04. 
The high-density matter with $2.5 \lesssim\log \xi \lesssim 5.5$
is blown away via the line force,
producing the funnel-shaped disk wind with an opening angle of 
$\sim 78^\circ$ (see Figure \ref{Fig1}).
In the Appendix \ref{App1}, we explain the launching mechanism 
and the dynamics of the wind in detail.

\begin{figure}
  \begin{center}
    \FigureFile(77mm,77mm){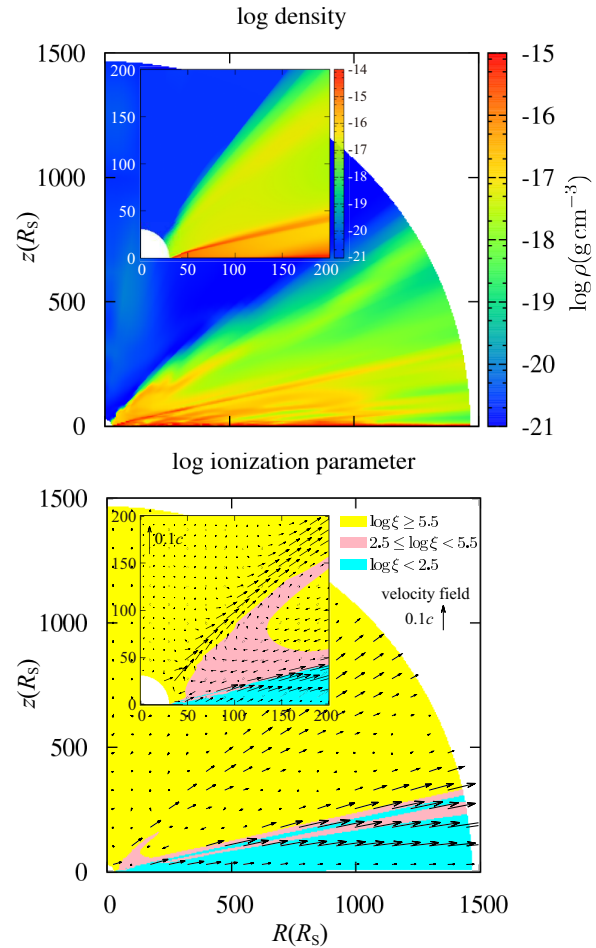}
  \end{center}
  \caption{Time averaged density map (top panel) and the ionization parameter map (bottom panel) for $\varepsilon=0.5$, $M_{\rm BH}=10^8 M_{\odot}$, $f_{\rm X}=0.1$, and $\rho_0=10^{-9}{\rm g\,cm^{-3}}$.
In the bottom panel, the yellow, pink, and cyan show the regions in which the ionization parameters are
$\log \xi \geq 5.5$, $2.5 \leq \log \xi < 5.5$, and $\log \xi<2.5$.
The foreground vectors show the velocity field.
The $z=0$ plain corresponds to the accretion disk surface whose height is $z_0=2R_{\rm S}$.
The $z$-axis is equal to the rotational axis of the accretion disk.
}\label{Fig1}
\end{figure}

\subsection{Comparison with observational features of UFOs}
\label{UFO}
Here, we concentrate on the comparison between 
the results of our simulations and the observations of UFOs.
Specifically, 
we investigate the ionization parameter, the outward velocity, and the column density along the line of sight (from an observer to the origin of the coordinate) based on our results. Then, we calculate the probability of detecting UFOs (UFO probability).
The definition of UFOs is that the blueshift of the absorption lines of FeX\hspace{-.1em}X\hspace{-.1em}V and/or FeX\hspace{-.1em}X\hspace{-.1em}V\hspace{-.1em}I is larger than $10,000\,{\rm km\,s^{-1}}$. Based on \citet{Tombesi11}, the ionization parameter spans $2.5 \lesssim\log \xi \lesssim 5.5$, and the column density is distributed in $22 \lesssim\log N_{\rm H} \lesssim 24$.
Thus, we recognize the UFO features to be detected
if the following conditions (A) and (B) are satisfied.
(A) the outward velocity of the matter with $ 2.5\leq \log\xi<5.5$ 
exceeds $10,000\,\rm{km\,s}^{-1}$.
(B) the column density of the matter with $2.5\leq \log\xi<5.5$ 
is larger than $10^{22}\,\rm{cm^{-2}}$.
We evaluate these conditions along the line of sight with the viewing angles between $\theta=0^{\circ}$ and $90^{\circ}$, and 
calculate the solid angle, $\Omega_{\rm UFO}$, that the UFOs could be observed.
We estimate the UFO probability as $\Omega_{\rm UFO}/4\pi$.
%

Figure \ref{Fig1} shows the time averaged wind structure 
of the fiducial model in $R$--$z$ plane,
where $z$-axis corresponds to the rotation axis of the disk
and $R$ is the distance from the rotation axis.
%
The top panel shows the density map.
The bottom panel shows the ionization parameter map
overlaid with the velocity vectors,
where the ionization parameter 
is presented by dividing into three levels;
the yellow ($\log \xi \geq 5.5$), 
pink ($2.5 \leq \log \xi < 5.5$), 
and cyan ($\log \xi<2.5$).
The extended view of the inner region 
($200R_{\rm S}\times 200R_{\rm S}$) 
is shown in the both panels.
%
The moderately ionized region ($2.5\leq \log\xi<5.5$) 
roughly corresponds to the mainstream of the wind described 
by the orange slimline structure 
(see the extended view of the top panel),
which is sandwiched by the low-ionization region ($\log\xi<2.5$)
and the fully-ionized region
($\log \xi \geq 5.5$). 
Along the mainstream, the dense matter moves outward with large
velocity, $\sim 0.1c$.
Thus, if this main stream gets across the line of sight, 
the blueshifted absorption features due to moderately ionized ions are expected. 

\begin{figure}
  \begin{center}
    \FigureFile(77mm,77mm){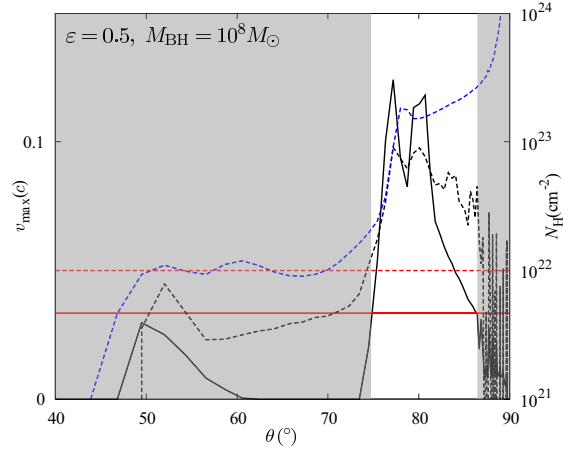}
  \end{center}
  \caption{Column density (dashed blue line), 
column density of the matter with $2.5\leq \log\xi<5.5$ (dashed line), 
and maximum velocity of the matter with $ 2.5\leq \log\xi<5.5$ (solid line) 
based on the time averaged results.
The abscissa is the viewing angle. The ordinate on the left-hand side shows the value of the solid line. The ordinate on the right-hand side shows the value of the dashed line.
The red solid line and the red dashed line show the lines of $v_{\rm max}=10,000\,{\rm km \, s^{-1}}$ and $N_{\rm H}=10^{22}\,{\rm cm^{-2}}$.
}\label{Fig2}
\end{figure}

%
Figure \ref{Fig2} 
shows the column density (dashed blue line),
the column density of the matter with $2.5\leq \log\xi<5.5$ (dashed black line), 
and the maximum velocity of the matter 
with $ 2.5\leq \log\xi<5.5$ (solid black line)
based on the time averaged results.
The abscissa is the viewing angle. The ordinate on the left-hand side shows the value of the solid line. The ordinate on the right-hand side shows the value of the dashed lines.
The red solid line and the red dashed line indicate $v_{\rm max}=10,000\,{\rm km \, s^{-1}}$ and $N_{\rm H}=10^{22}\,{\rm cm^{-2}}$.
The conditions (A) and (B) are satisfied 
at the unshaded region
($75^{\circ}\lesssim\theta \lesssim 86^{\circ}$), where 
the solid black line is above the red solid line and 
the dashed black line is above the red dashed line.
The resulting UFO probability is 20\%.



In the region of $\theta \lesssim 50^{\circ}$, 
there is no dense and fast wind, 
so neither of the conditions (A) nor (B) is satisfied. 
In the region of $50^{\circ} \lesssim \theta \lesssim 75^{\circ}$, 
the column density is around $10^{22}\,{\rm cm^{-2}}$, 
but a large portion of the matter is highly ionized 
($\log \xi \geq 5.5$).
In this region, 
the density is relatively small and the matter is not shielded 
from the X-ray irradiation.
As a consequence, the column density of the matter with $2.5\leq \log\xi<5.5$ 
is smaller than $10^{22}\,{\rm cm^{-2}}$.
The outflow velocity of the matter with $2.5\leq \log\xi<5.5$ is 
small due to the weak line force.

Near the the disk surface, $\theta \gtrsim 86^{\circ}$, 
the radial velocity is too small 
to satisfy the condition (A),
since the radial component of the radiation from the disk 
is roughly proportional to $\cos\theta$ and 
is attenuated by the high-density matter around the base of the mainstream.
The X-ray is also diluted around the base of the mainstream,
a large fraction of the matter is in a low-ionization state, 
$\log \xi < 2.5$ (see cyan in the bottom panel of Figure 1).
Thus, the condition (B) is not satisfied though 
the column density is $10^{23}$--$10^{24}\,{\rm cm^{-2}}$.
We do not expect to observe the UFOs if we observe the accretion disk edge-on.
%
Here, we notice that the X-ray spectrum that \citet{Tombesi11} assumes 
for estimation of the ionization parameter of UFOs 
(the power-law with a photon index of 2 
in the energy range between $13.6\,{\rm eV}$ and $13.6\,{\rm keV}$) 
is different from that implicitly assumed in this simulation 
(see section \ref{SED} for detailed discussions).


\subsection{Eddington ratio dependence}
\label{Edd}
Here, we discuss the Eddington ratio dependence of the UFO probability.
We calculate the disk wind for a wide range of the Eddington ratio 
while the black hole mass is kept constant here, $M_{\rm BH}=10^8 M_{\odot}$.
We find that the UFO probability is around 20\% and less sensitive to 
the Eddington ratio in the range of $\varepsilon=0.1$--$0.7$.
The time averaged UFO probabilities are 17\%, 19\%, 20\%, and 27\% 
for $\varepsilon=0.1,$ $0.3,$ $0.5$ and $0.7$. 
%
To understand the reason why the UFO probability 
has roughly the same value for the different Eddington ratios, 
we compare the wind structure of $\varepsilon=0.1$ 
to that of $\varepsilon=0.5$.

\begin{figure}
  \begin{center}
    \FigureFile(77mm,77mm){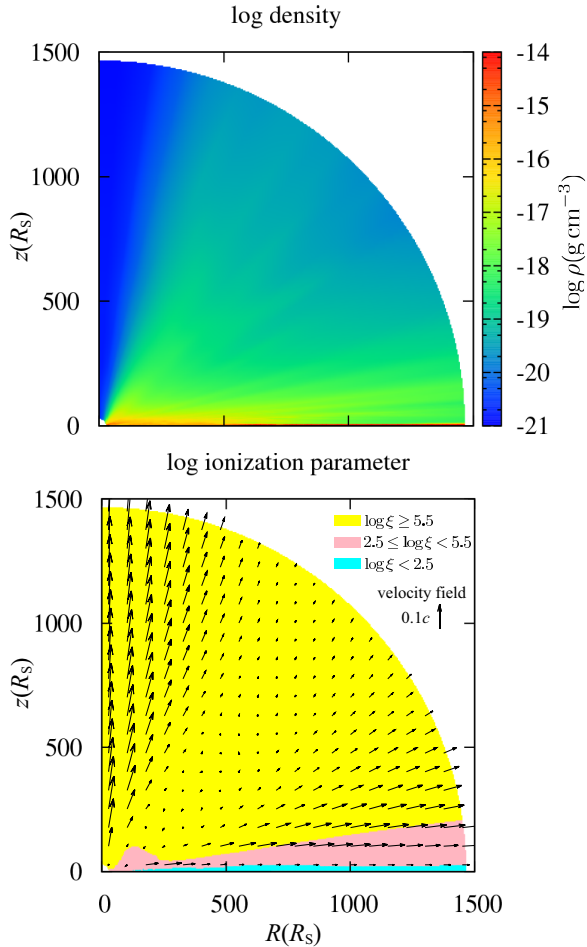}
  \end{center}
  \caption{
Same as Figure \ref{Fig1}, 
but for $\varepsilon=0.1$. 
}\label{Fig2-0}
\end{figure}

Figure \ref{Fig2-0} is the same as Figure \ref{Fig1}, 
but for $\varepsilon=0.1$.
In almost whole region,
the density is smaller than that for $\varepsilon=0.5$ 
 (see the top panel). 
This is because
the radiation force is not large enough to accelerate a large amount of the matter.
Thus, the region with $\log \xi \geq 5.5$ 
occupies the most volume
(see the bottom panel),
since the less dense matter is highly ionized by the X-ray.
The relatively high-density matter exists at only the region of 
$R\lesssim 100R_{\rm S}$ and $\theta \gtrsim 70^\circ$
(orange).
This region has the ionization parameter 
of $2.5 \leq \log \xi < 5.5$.
The density of this region is too small to be $\log \xi <2.5$,
but this region works to shield the X-ray.
Thus, behind this region ($R\gtrsim 100R_{\rm S}$), 
the ionization parameters are $2.5 \leq \log \xi < 5.5$
and $\log \xi < 2.5$.

%

\begin{figure}
  \begin{center}
    \FigureFile(77mm,77mm){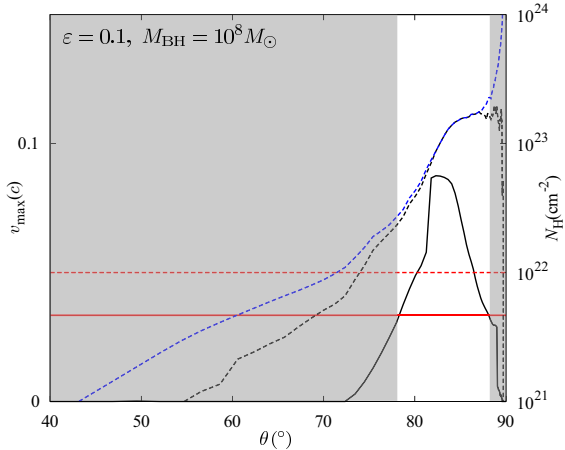}
  \end{center}
  \caption{Same as the top panel of Figure \ref{Fig2}, 
but for $\varepsilon=0.1$.
}\label{Fig2-e1}
\end{figure}

Figure \ref{Fig2-e1} is the same as Figure \ref{Fig2}, but for $\varepsilon=0.1$. 
We find that the angular range in which the condition (A) is satisfied 
is restricted to $\theta \sim 78$--$88^\circ$,
although the condition (B) is satisfied at $\theta\gtrsim 73^\circ$.
Thus,
as compared with that of $\varepsilon=0.5$,
the unshaded region slightly shifts to large $\theta$, 
but the UFO probability is almost the same.
%
In the unshaded region,
we find that the column density is smaller for $\varepsilon=0.1$
than for $\varepsilon=0.5$, 
however the column density of the matter with $2.5\leq \log\xi<5.5$ 
is larger for $\varepsilon=0.1$ than for $\varepsilon=0.5$.
This is because
the most of matter is in moderate-ionization state for $\varepsilon=0.1$
(see the bottom panel of Figure \ref{Fig2-0}).
We find the column density is equal to that of the matter with $2.5\leq \log\xi<5.5$.
In contrast, for the case with $\varepsilon=0.5$,
a substantial proportion of the matter 
is in low-ionization state ($\log\xi<2.5$)
in the unshaded region of Figure 2.
The Figure 4 also shows that 
the maximum velocity slightly decreases with a 
decline of the Eddington ratio because of the reduced radiation force.
For $\varepsilon=0.1$, the maximum velocity is smaller than $0.1c$.


%

In the case of $\varepsilon=0.01$,
the disk wind is not launched, 
since the radiation force is too small to accelerate the disk wind.
Thus, the UFO probability is zero. 

\subsection{Black hole mass dependence}
\label{BH}
Next we examine how the UFO probability depends on the black hole mass
in a range between 
$M_{\rm BH}=10^6 M_{\odot}$ and $10^9 M_{\odot}$, 
for a fixed Eddington ratio, $\varepsilon=0.5$.
We find that the funnel-shaped disk winds appear for all cases,
and the UFO probability
is almost constant or slightly decreases 
with the increase of the black hole mass,
28\%, 22\%, 20\%, and 13\% for 
$M_{\rm BH}=10^6 M_{\odot}$, $10^7 M_{\odot}$, $10^8 M_{\odot}$, and $10^9 M_{\odot}$.

\begin{figure}
  \begin{center}
    \FigureFile(77mm,77mm){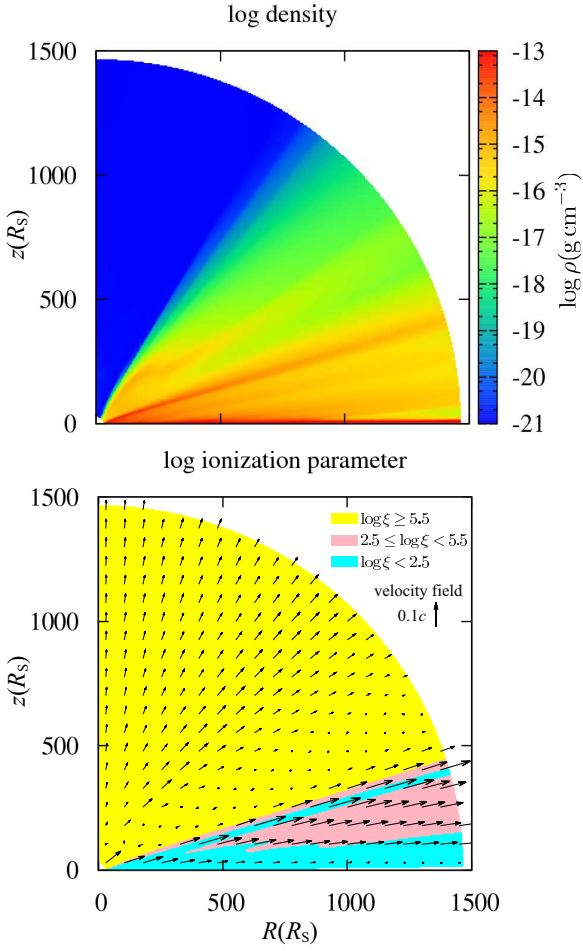}
  \end{center}
  \caption{
Same as Figure \ref{Fig1}, 
but for $M_{\rm BH}=10^6 M_{\odot}$. 
}
\label{Fig4-1}
\end{figure}

\begin{figure}
  \begin{center}
    \FigureFile(77mm,77mm){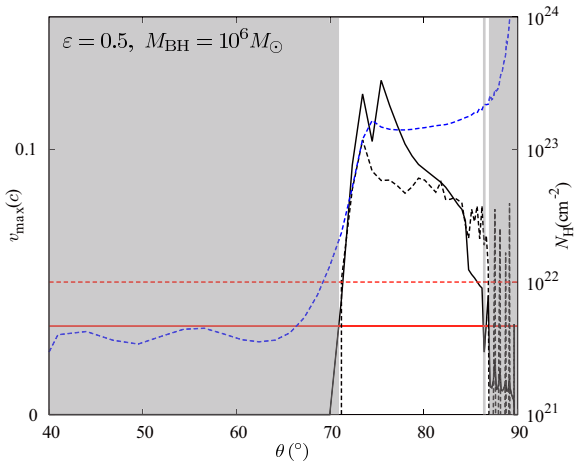}
  \end{center}
  \caption{Same as the top panel of Figure \ref{Fig2}, 
but for $M_{\rm BH}=10^6 M_{\odot}$.
}\label{Fig4-2}
\end{figure}

Figures \ref{Fig4-1} and \ref{Fig4-2} are the same as 
Figures \ref{Fig1} and \ref{Fig2},
but for $M_{\rm BH}=10^6 M_{\odot}$. 
Figure \ref{Fig4-2} indicates that 
the range of the polar angle in which the UFOs are observed 
(the unshaded region in the plot) is similar to that of 
$M_{\rm BH}=10^8 M_{\odot}$.
Since the density of the wind
is roughly two orders of 
magnitude larger than that for $M_{\rm BH}=10^8 M_{\odot}$
($\rho \propto M_{\rm BH}^{-1}$),
the column density ($\propto \rho r$)
in the unshaded region 
is insensitive to $M_{\rm BH}$.
We also find that the velocity and the column density
of the matter with $2.5\leq \log\xi<5.5$
($\gtrsim 0.1c$ and $\sim 10^{23}\,{\rm cm^{-2}}$)
are insensitive to the black hole mass,
leading to insensitivity of the UFO probability.

Such a insensitivity is understood as follows.
At a given Eddington ratio,
the force multiplier, $M(t, \xi)$, 
required to successfully launch the wind
does not depend on the black hole mass.
The wind velocity 
roughly corresponds to the escape velocity at the launching region
of $R\sim 30R_{\rm S}$ ($v_r \propto M_{\rm BH}^0$),
so that $dv_r/dr$ is proportional to $r^{-1}\propto M_{\rm BH}^{-1}$,
leading to $t \propto \rho |dv_r/dr|^{-1} \propto \rho M_{\rm BH}$.
On the other hand,
we have $\xi\propto L_{\rm X}/\rho r^2 \propto \rho^{-1} M_{\rm BH}^{-1}$
because of $L_{\rm X}\propto L_{\rm Edd} \propto M_{\rm BH}$.
Hence, the force multiplier does not depend on the black hole mass
when $\rho \propto M_{\rm BH}^{-1}$,
and the wind structure as well as the UFO probability 
is insensitive to the black hole mass.

\begin{figure}
  \begin{center}
    \FigureFile(77mm,77mm){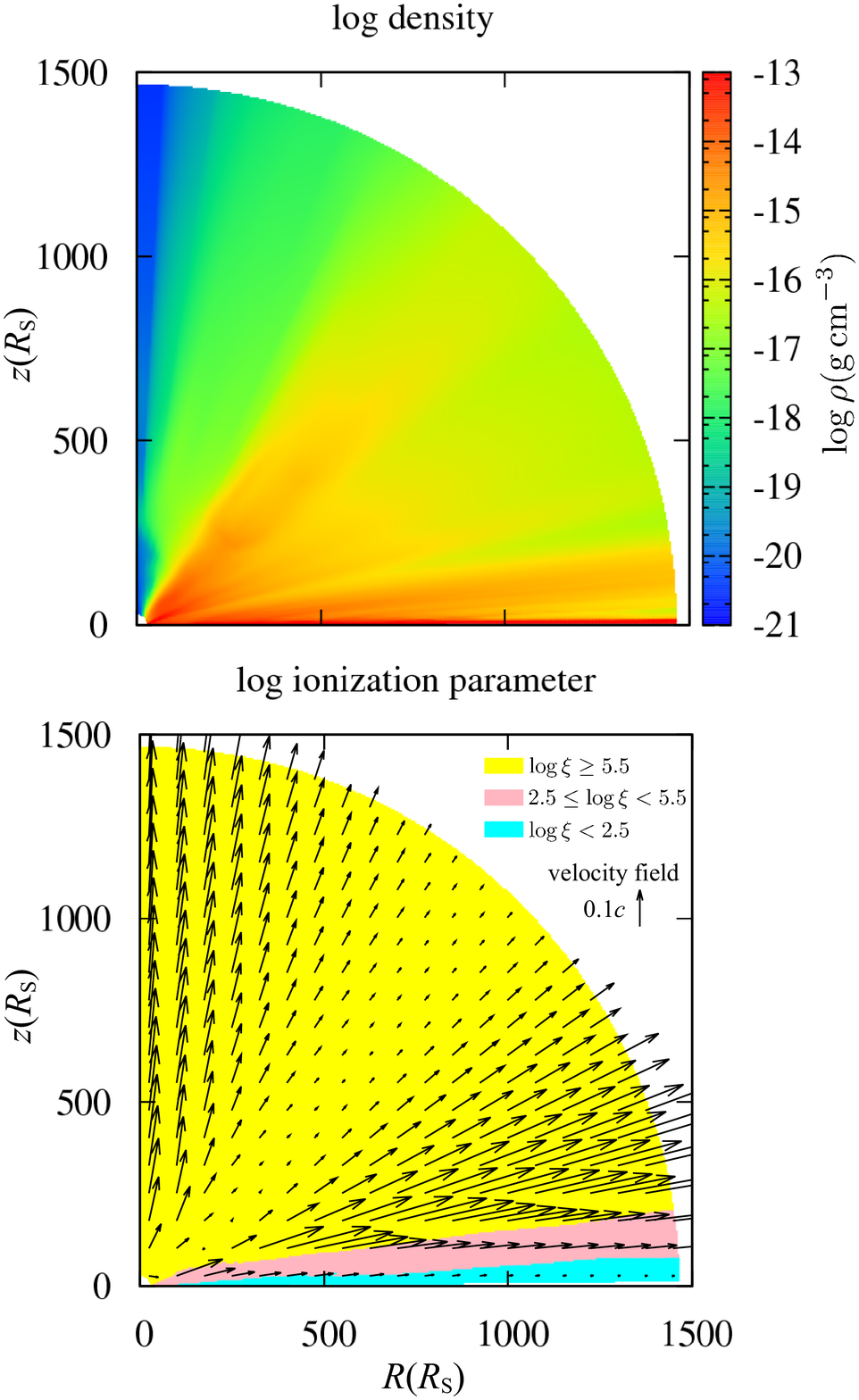}
  \end{center}
  \caption{
Same as Figure \ref{Fig4-1},
but for $f_{\rm X}=1$. 
}\label{Fig5-1}
\end{figure}

\begin{figure}
  \begin{center}
    \FigureFile(77mm,77mm){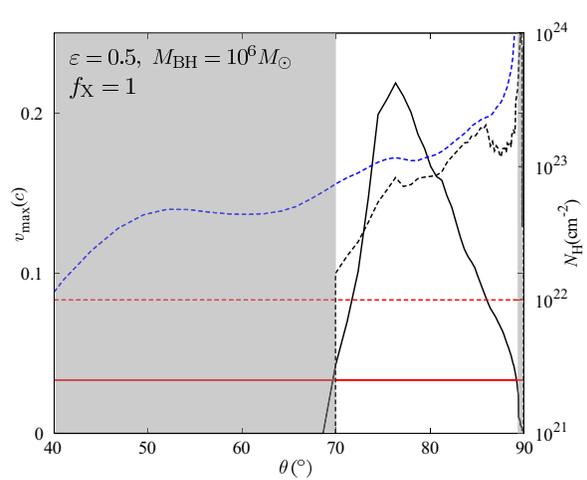}
  \end{center}
  \caption{Same as Figure \ref{Fig4-2}, 
but for $f_{\rm X}=1$. 
}\label{Fig5-2}
\end{figure}

\subsection{X-ray luminosity dependence}
\label{XUV}
In order to understand the effect of the X-ray irradiation, 
we calculate the disk wind under the large X-ray luminosity,
$f_{\rm X}=1$ (i.e., $L_{\rm X}=0.5L_{\rm Edd}$).
The disk luminosity and the black hole mass are set to be
$L_{\rm D}=0.5L_{\rm Edd}$ and $M_{\rm BH}=10^6 M_{\odot}$.
In this case, since the X-ray luminosity is comparable to the disk luminosity, 
we should include the radiation force by the X-ray in reality.
However, in order to investigate the effect of the ionization by the X-ray irradiation,
we neglect the X-ray radiation force as the case of $f_{\rm X}=0.1$.
We find that 
the resultant UFO probability is 34\% for $f_{\rm X}=1$, 
which is comparable or slightly larger than that for $f_{\rm X}=0.1$, 28\%.

Figure \ref{Fig5-1} 
is the same as Figure \ref{Fig4-1},
but for $f_{\rm X}=1$.
We find that 
the moderate- and low-ionization regions
exist only near the disk surface.
This is because the 
effect of the photoionization via the X-ray is enhanced.
However, the moderately ionized matter near the launching region,
$R\sim 30R_{\rm S}$, obscures 
the X-ray source with the angular range of $\theta \gtrsim 70^\circ$.
Thus, the UFO probability does not decrease 
even when we employ $f_{\rm X}=1$.
This is clearly shown in Figure \ref{Fig5-2},
which is the same as Figure \ref{Fig4-2}, but for $f_{\rm X}=1$.
In the angular range of $\theta \gtrsim 70^{\circ}$,
the column density with $2.5\leq \log \xi<5.5$
exceeds $10^{22}\,{\rm cm^{-2}}$ so that 
the condition (B) is satisfied.
Since the condition (A) is also satisfied in this range,
the UFO features are observed
for the observer with $\theta \gtrsim 70^\circ$.
Such result is almost the same as the result of $f_{\rm X}=0.1$. 

Here, we note that
the high-density region appears
in the direction of $\theta \sim 45$--$70^\circ$
as shown in the top panel in Figure \ref{Fig5-1}.
However, this region does not raise the UFO probability,
since both the ionization state and the velocity 
disagree with the observations
(see the bottom panel in Figure \ref{Fig5-1} and Figure \ref{Fig5-2}).
In the case of $f_X=1$, 
the line force works effectively only at the region that 
the X-ray is heavily obscured.
In this region, the radial component of the disk flux is also
diluted so that a part of the matter is blown away
relatively upward ($\theta \sim 45$--$70^\circ$) via the line force. 

\section{Discussion}
\label{discussions}




\subsection{UFO probability, kinetic luminosity, and outflow rate in observations}
Our present simulations of the line-driven disk winds 
are roughly consistent with the observations of the UFOs.
Our model shows that the UFO probability is 13--28\%
in a wide parameter range of the Eddington ratio and the black hole mass
($\varepsilon \geq 0.1$ and $M_{\rm BH}=10^{6-9}M_\odot$).
We also found that this probability is insensitive to 
the ratio of the X-ray luminosity to the disk luminosity
and the density of the disk surface 
(see Appendix \ref{App2}).
The resulting probability of 13--28\%
is roughly comparable to the estimation by \citet{Tombesi11}.
They reported the UFO probability to be $\sim $40\%
based on the frequency of detecting UFOs in the Seyfert galaxies,
although the number of UFO samples is still small 
and the parameter dependence of UFOs is still unknown.
As shown in section \ref{XUV},
the wind is successfully launched 
when the black hole mass is relatively small ($M_{\rm BH}=10^6 M_{\odot}$)
and the X-ray luminosity is large ($L_{\rm X}=0.5L_{\rm Edd}$).
This implies that the UFO features could appear in the narrow line Seyfert 1
galaxies (NLS1s).
In fact, the UFO 
is detected in the NLS1, PG1211+143
\citep{Pounds03,Tombesi11}. 
According to the unified model of AGNs, 
the torus is thought to be located far outside of the accretion disk 
and obscures the nucleus. 
If the torus is Compton thin or clumpy, 
UFO features can be observed even if the torus interrupts the line of sight of the observer. 
However, if the rotational axis of the accretion disk is aligned with that of the torus, 
and if the torus is optically thick enough to dilute the UFO features, 
the range of viewing angles in which the UFOs are observed decreases, 
reducing the UFO probability. 
If the outflow is coexistence with such type of AGNs, we cannot detect the absorption lines from near equatorial direction.
Then, the UFO probability that we estimated is an upper limit. 
On the other hand, if the rotational axis of the Compton-thick torus is misaligned with that of the accretion disk, some of AGNs, which do not show the UFO features intrinsically, are not detected via the obscuration. This leads to increase of the UFO probability.

%
The mass outflow rate and the kinetic luminosity 
of the line-driven disk winds in the present simulations
are also consistent with the observations of the UFOs.
We evaluate the mass outflow rate 
as
$\dot M_{\rm out}=4\pi r^2 
\int^{89^{\circ}}_0 \rho v_r \sin\theta d \theta$
at the outer boundary of the calculation box, $r=r_{\rm o}(=1500R_{\rm S})$.
\footnote[1]{
We exclude the mass flux of $89$--$90^\circ$ from the integration 
in order to eliminate the influence of the boundary condition 
of the disk surface. 
In the region of $r\sim r_{\rm o}$ and $\theta\gtrsim 89^\circ$, 
the density of the outflow is large in consequence of 
the boundary condition 
that the density is fixed to be $\rho_0$ at the $\theta=90^\circ$ plane. 
}
%
We find $\dot M_{\rm out}=0.03$--$0.05 \dot M_{\rm acc}$
for a wide range of $\varepsilon$ and $M_{\rm BH}$, 
where $\dot M_{\rm acc}$ is the accretion rate 
derived from $\dot M_{\rm acc} = \varepsilon L_{\rm Edd}/\eta c^2$ 
with $\eta=0.06$.
This is consistent with that estimated in X-ray observations, 
i.e. $\dot M_{\rm out}\gtrsim 0.05$--$0.1 \dot{M}_{\rm acc}$ \citep{Tombesi12}
although the opening angle of the outflows is observationally less certain.

The kinetic luminosity of the outflows is evaluated by
$L_{\rm kin}=2\pi r^2 
\int^{89^{\circ}}_0 \rho v_r^3 \sin\theta d \theta$
at $r=r_{\rm o}$,
which is
0.5\%--1\% 
of the disk luminosity for $\varepsilon \geq 0.1$ and
$M_{\rm BH}\geq 10^7 M_{\odot}$. 
For the cases of 
$(\varepsilon, M_{\rm BH})=(0.1, 10^8M_\odot)$
and $(0.5, 10^6M_\odot)$,
the kinetic luminosities are 0.1\% and 0.3\% of the disk luminosity.
\citet{Tombesi12} have reported that the kinetic luminosity
is about $\gtrsim 0.3$\% of the bolometric luminosity ($L_{\rm bol}$),
by using the bolometric correction, 
$L_{\rm bol}\sim 30L_{\rm 2-10}$ with $L_{\rm 2-10}$ being
the $2$--$10\,\rm{keV}$ luminosity.
Since the bolometric luminosity is almost the same as the disk luminosity,
($L_{\rm bol} \sim L_{\rm D}$),
our present result agrees with the X-ray observations of UFOs.

In addition, in the model presented by \citet{Hopkins10}, 
the energy outflow rate is required to be larger than 0.5\% 
of the bolometric luminosity to have the significant 
feedback on the surrounding environment.
Thus, our present simulations imply that 
the line-driven disk wind 
could contribute to the feedback on the host galaxies. 



Our present study predicts
a fraction of quasars show the UFO features,
since the disk wind is launched from the disks
around supermassive black holes, $M_{\rm BH}\sim 10^9M_{\odot}$,
as long as the Eddington ratio is larger than 0.1.
However, the disk winds do not appear 
for the less luminous sources with $\varepsilon=0.01$.
This indicates that the UFOs are not detected 
in the less luminous AGNs.
As we have already mentioned above,
NLS1s are the candidate for exhibiting the UFO features.
This will be verified 
statistically larger samples of UFOs in future observations.

\subsection{Spectral energy distributions}
\label{SED}
The line force depends on the spectral energy distributions (SEDs).
The force multiplier presented by SK90, 
which is used for the present simulations, 
is produced by assuming that 
the $10\,{\rm keV}$ bremsstrahlung SED determines the ionization states of the various metals 
and the radiation from a star with an effective temperature of $25,000\,{\rm K}$
contributes to the line force. 
However, the typical SED of the AGNs is the multicolor blackbody superposed with the power-law. 

In order to assess the difference between the force multiplier by SK90 and that for the AGN SED, 
we calculate the radiation transfer of the incident AGN SED 
and estimate the effective force multiplier from the attenuation of the radiative flux. 
In this calculation, 
we use a spectral synthesis code, Cloudy (version 13.03, described by \cite{Cloudy}), 
and adopt the density and the velocity gradient ($\rho=2.04\times 10^{-16}\,{\rm g\,cm^{-3}}$, 
$\Delta v_r/\Delta r=0.0121c/R_{\rm S}$) 
of a grid cell ($\Delta r=0.616R_{\rm S}$) 
at around the wind base of the main stream ($r=31.8R_{\rm S}$, $\theta=89.3^{\circ}$), 
where the line force strongly accelerates the matter and drives the wind. 
The metal abundance is assumed to be the solar abundance. 

The incident SED consists of the multicolor disk blackbody 
and the power-law. 
Here, the multicolor disk blackbody is 
the superposition of the blackbody radiation from the standard type disk, 
of which the effective temperature is described as
$1.8\times 10^5(r/3R_{\rm S})^{-3/4}\,{\rm K}$ 
(see also the equation [\ref{disktemp}]).
We do not take into account the radiation 
from the region of $T_{\rm eff}<3\times 10^3\,{\rm K}$ ($r>2315R_{\rm S}$) 
because the effective temperature is too cold 
to emit the photons largely contributing to the line force. 
The SED of the multicolor disk blackbody has a peak at around $30\,{\rm eV}$. 
The photon index of the power-law spectrum in the energy range of 
$0.1$--$10^5\,{\rm keV}$ is $2.1$. 
The flux of the power-law is set to be 10\% of the multicolor disk blackbody 
radiation. 
We adjust the intensity of the incident radiation 
as the ionization parameter becomes nearly equal to 
that of the hydrodynamic simulations ($\xi=0.773$) 
in order to focus on the difference of the SED shape.
We recognize the ratio of the attenuation via the metal absorption 
and that due to the electron scattering as the force multiplier for the AGN SED.

As a result, we find that the force multiplier for the AGN SED is about 3.6 times lager than that of SK90. 
This is because the radiation in a band between $13.6\,{\rm eV}$ and $1\,{\rm keV}$, 
which is stronger in the AGN SED than in the SED employed in SK90, 
largely contributes to the line force. 
Since the radiation in this band is effectively absorbed, 
such an enhanced line force would not realized except at the very vicinity of the radiation source. 
That is, a factor 3.6 is probably an upper limit. 
Indeed, the optical thickness from the disk surface to the selected grid cell is larger than 10. 


To research the effects of the enhancement of the force multiplier, 
we run a test model, in which the force multiplier is 3.6 times larger than that of SK90 for all the grid cells. 
As a result, we find that the maximum outflow velocity is around $0.3c$, 
which is three times larger than the original result ($\sim 0.1c$). 
The UFO probability is 41\%, 
which is about twice as large as the original UFO probability (20\%). 
This is because the enhanced line force accelerates 
the wind more upward and the solid angle of the wind increases.
The velocity of $\sim 0.3c$ is not inconsistent with observations 
and the probability of $\sim 40\%$ fits the estimation by \citet{Tombesi11}.
Hence, our disk wind model is consistent with the observed UFO features 
even if the force multiplier is revised for the AGN SED. 


Here we note that, 
even if the force multiplier for the AGN SED can be proposed, 
the obtained line force is not rigorous. 
The spectral shape changes due to the gas-radiation interaction 
as the radiation passes through the medium. 
However, this effect is not taken into account in the force multiplier, 
which is produced based on the fixed spectral shape. 
The spectrum for each place is obtained by solving the three-dimensional, 
multi-frequency radiation transfer calculations. 
\citet{HP14} performed the radiation transfer calculations 
by employing the density structure of the snapshot obtained by PK04. 
Although they reported the distribution of the ionization parameter in the case of the AGN SED, 
their post-process calculation could not reveal the wind dynamics. 
The hydrodynamic simulations coupled with the three-dimensional, 
multi-frequency radiation transfer calculations are for further study.

\citet{Tombesi11} has reported that the observed absorption lines of 
FeX\hspace{-.1em}X\hspace{-.1em}V 
and/or FeX\hspace{-.1em}X\hspace{-.1em}V\hspace{-.1em}I
are reproduced 
by the outflowing matter with $2.5 \lesssim\log \xi \lesssim 5.5$, 
by employing the power-law X-ray spectrum with the photon index,
$\Gamma$, of 2, 
and by considering that the photons in a band 
from $13.6\,{\rm eV}$ to $13.6\,{\rm keV}$ contribute to 
the ionization of the metals. 
Since our simulations employ the force multiplier proposed by SK90, 
the ionization parameter in the present work is evaluated 
by assuming the $10\,{\rm keV}$ bremsstrahlung by implication.


To investigate the difference of the absorption lines
due to the difference in the employed SED, 
we use Cloudy and calculate equivalent widths of
FeX\hspace{-.1em}X\hspace{-.1em}V 
and FeX\hspace{-.1em}X\hspace{-.1em}V\hspace{-.1em}I 
for the $10\,{\rm keV}$ bremsstrahlung and the power-law with $\Gamma=2$. 
We focus on a grid cell of $\log \xi\sim 4.0$ 
located at $r=95.9R_{\rm S}$ and $\theta=78.0^{\circ}$. 
The density is $\rho=3.8 \times 10^{-17}\,{\rm g\,cm^{-3}}$, 
the grid size is $\Delta r=7.6R_{\rm S}$, and the outflow velocity is $v_r=0.028c$. 
This grid cell includes the angle
in which the conditions for detecting UFO 
(see section \ref{UFO}) are satisfied. 
We adjust the intensity of the incident radiation 
as the ionization parameter is consistent with 
the result of the hydrodynamic simulations. 
Thereby, we can focus on the difference of the spectral shape.

Resultingly, 
the absorption lines of 
FeX\hspace{-.1em}X\hspace{-.1em}V 
and FeX\hspace{-.1em}X\hspace{-.1em}V\hspace{-.1em}I
appear for both SEDs, 
and the equivalent width is not significantly different.
We find that the equivalent widths of
FeX\hspace{-.1em}X\hspace{-.1em}V 
and FeX\hspace{-.1em}X\hspace{-.1em}V\hspace{-.1em}I
for the 10keV bremsstrahlung are $\sim 0.35$ times and $\sim 0.65$ times of those 
for the power-law. 
The ratio of the equivalent widths of 
FeX\hspace{-.1em}X\hspace{-.1em}V 
and FeX\hspace{-.1em}X\hspace{-.1em}V\hspace{-.1em}I
is $\sim 0.52$ times of that for the power-law.

These results imply that 
we should take into account the difference in the SEDs of the incident radiation 
when we try to compare absorption features 
between the models and observations in great detail.
However, this is not serious for the present study, 
which is roughly researching $M_{\rm BH}$-dependence and $\varepsilon$-dependence 
of the UFO probability.
The theoretical spectral synthesis adopting the realistic SED is necessary 
to discuss about the absorption features 
\citep{Schurch09,Sim10}. 
We will report the simulated absorption features 
and comparison with the observations of UFOs in a forthcoming paper.

\subsection{Future works}
In our simulations, 
we simply assume that the radiation from the accretion disk contributes to 
the line force and the ionization parameter is determined by 
the radiation from the vicinity of the black hole 
(approximated as a point source). 
To study the influence of the ionization photons from the disk, 
we calculate the ionization parameter 
by considering the radiation of $>13.6\,{\rm eV}$ 
from the accretion disk as follows,
$\xi_{\rm D}=4\pi F_{\rm X}/n + (4\pi/n)\{(F_{r, {\rm ion,thin}}e^{-\tau_{\rm X}})^2
+ F_{\theta,{\rm ion,thin}}^2\}^{0.5}$,
%
where the first term indicates the contribution from the central point source 
and the second term is the contribution from the disk radiation. 
In the second term, $F_{r, {\rm ion,thin}}$ and $ F_{\theta,{\rm ion,thin}}$ are
the $r$-component and the $\theta$-component of the flux 
above $13.6\,{\rm eV}$ (or $0.1\,{\rm keV}$) emitted from the disk 
in the optically thin media. 
In the same way as the estimation of the disk flux, 
the $r$-component of the flux is diluted in accordance with the optical depth 
measured from the center of the coordinate, $\tau_{\rm X}$ (see the equation [\ref{tau_iX}]).

%
As a result, for our fiducial model, 
highly ionized region ($\log \xi \geq 5.5$) largely expands 
and moderate- and low- ionization regions ($2.5 \leq \log \xi < 5.5$ and $\log \xi<2.5$)
appear only near the disk surface. 
Thus, the UFO probability is reduced to $\sim 8\%$, 
which is about a half of our estimation described in section \ref{UFO}. 
Since the low-ionization region ($\log \xi<2.5$), 
in which the acceleration due to the line force is effective, 
becomes small 
from $r\gtrsim 30R_{\rm S}$ and $\theta\gtrsim 78^{\circ}$ (see Figure \ref{Fig1}) 
to $r\gtrsim 140R_{\rm S}$ and $\theta\gtrsim 88^{\circ}$,
the wind power might decrease.

%
However, this ionization parameter would be overestimated, 
because the attenuation of the $\theta$-component of the radiation flux is neglected. 
The photons of which energy is $\gtrsim 13.6\,{\rm eV}$ are largely diluted 
in the high-density region around the disk surface in reality, 
so that the high-ionization region does not expand so much. 
Indeed, the resulting distribution of the ionization parameter 
is almost the same as Figure \ref{Fig1}, 
if we estimate the ionization parameter 
by excluding the radiation of $13.6\,{\rm eV}$-$0.1\,{\rm keV}$. 
This is the case, both the wind dynamics 
as well as the UFO probability would not change so much.




We assume axial symmetry in the present simulations, 
but the three-dimensional simulations would be an important future work.
Recently, the time variations of the absorption lines are discovered
(e.g., \cite{Misawa07,Cap13,Tombesi12b}).
This suggests that the structure of the disk wind 
changes in time and/or the wind has a non-axial symmetric structure.
In the one-dimensional as well as the two-dimensional calculations, 
the density fluctuations are reported (\cite{OP99}, PSK00).
In order to reveal 
the cause of the time variability, 
we should perform the time dependent three-dimensional simulations.

Although our simulations investigate the disk wind 
without solving the structure of the accretion disk,
it is a important work to calculate the wind and the disk self-consistently.
We assume that
the geometrically thin and optically thick standard disk 
lies just below our computational domain.
The $\theta=90^{\circ}$ plane is supposed to correspond to the disk surface,
and the density on the grids of $\theta=90^{\circ}$ does not change with time.
It implies that the mass and mass accretion rate of the disk 
does not decrease even if the disk wind is launched.
In addition, we treat the accretion disk as the external radiation source, 
and the photons are steadily emitted from 
the equatorial plane of the disk.
However, in reality, the disk emission
would change when the disk wind appears and the 
mass accretion decreases.
The disk and wind should be self-consistently treated 
by the multi-dimensional simulations.
Global two-dimensional radiation hydrodynamic/magnetohydrodynamic simulations
of the standard disk and the wind were performed 
by \citet{2006ApJ...640..923O, 2009PASJ...61L...7O, 2011ApJ...736....2O}.


\section{Conclusions}
\label{conclusions}
We performed two-dimensional radiation hydrodynamic simulations 
of line-driven disk winds in AGNs 
and compared our results to the observational features of UFOs.
As a consequence, 
we revealed that
the line force successfully launches the disk wind,
and the line-driven disk wind 
explains major properties of 
the UFOs.

For fiducial model
($\varepsilon=0.5$ and $M_{\rm BH}=10^8 M_{\odot}$), 
the funnel-shaped wind, of which the velocity is $\sim 0.1c$ and
the opening angle is $\sim 78^\circ$, is 
mostly
launched from the disk surface of $R\sim 30R_{\rm S}$.
The column density, the ionization state, and the velocity 
of the wind nicely agree with the observations of the UFOs
(blueshifted absorption features
of 
FeX\hspace{-.1em}X\hspace{-.1em}V and/or
FeX\hspace{-.1em}X\hspace{-.1em}V\hspace{-.1em}I),
by which the matter of $> 10^{22}\,\rm{cm^{-2}}$ 
is in the moderate-ionization state, $2.5\leq \log \xi < 5.5$,
and is ejected with velocity of $> 10,000\,\rm{km\,s}^{-1}$.
This result indicates that the UFO features are detected 
when we observe the nucleus from the viewing angles $\theta$,
between $75^{\circ}$ and $86^{\circ}$. 
The UFO features 
could be detected only
for the observer
with the large viewing angle, 
at least for 
the Eddington ratio is 
$\varepsilon=0.1$--$0.7$
and the black hole mass is $M_{\rm BH}=10^6$--$10^9M_\odot$.
The angular range of the viewing angle
is roughly independent of $\varepsilon$ and $M_{\rm BH}$
or slightly increases with an increase of $\varepsilon$
and a decrease of $M_{\rm BH}$.
Our simulations suggest that the UFOs are detected 
with a probability of 13--28\%,
if we observe the luminous AGNs ($\varepsilon\gtrsim 0.1$)
like Seyfert galaxies and quasars.
The probability is roughly consistent with 
that inferred from the X-ray observations of the UFOs, $\sim 40\%$
\citep{Tombesi11}.
The mass outflow rate as well as the kinetic luminosity of 
the wind also agree with the observations \citep{Tombesi12}.

Above results do not change significantly 
if we employ large X-ray luminosity,
suggesting
that the NLS1s are the candidates for exhibiting the UFO features.
In contrast, the less luminous AGNs does not show the UFOs,
since the wind is not launched for $\varepsilon \lesssim 0.01$.

\bigskip

We would like to thank Masayuki Umemura for useful discussions.
Numerical computations were carried out on Cray XC30 at the Center for Computational Astrophysics, CfCA, 
at the National Astronomical Observatory of Japan. 
M.N. is supported by the Research Fellowship from the Japan Society for the Promotion of Science (JSPS) 
for Young Scientists (25$\cdot$10444).
This work is supported in part
by Grants-in-Aid of the Ministry of Education,
Culture, Sports, Science and Technology (MEXT) 
(24740127, K.O.; 23540267, K.W.)
A part of this research has been funded by 
MEXT HPCI STRATEGIC PROGRAM and the Center for the Promotion of Integrated Sciences (CPIS) of Sokendai.


\appendix

\section{Dynamics of disk wind}
\label{App1}

In the extended view of the top panel in Figure \ref{Fig1}, 
We find streams of dense matter (slimline winds)
along the direction of $\theta\sim 78^{\circ}$ (orange).
This slimline wind, 
which possesses almost all the momentum of the outflow,
is the mainstream of the disk wind.
The matter of the mainstream
is lifted relatively upward around the wind base 
(launching region of the wind)
and turns in the radial direction,
producing the funnel-shaped wind. 
The reason why the slimline wind is kept narrow
is probably that the matter of the mainstream moves 
with the supersonic velocity.
Here, we discuss the acceleration mechanism of the matter 
around the wind base and the bending mechanism of the mainstream.



\begin{figure}
  \begin{center}
    \FigureFile(77mm,77mm){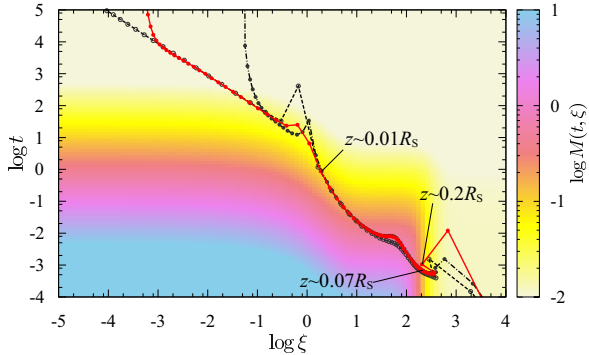}
  \end{center}
  \caption{
Force multiplier in the space of the ionization parameter and 
the local optical depth parameter.
%
The solid red line with red filled circles 
shows the location in $t$-$\xi$ space 
obtained by 
tracing the $R=30R_{\rm S}$ line from 
$z\sim 0$ to $z\sim 0.2$ ($\theta\sim 89.98^{\circ}$)
based on our results
for the fiducial model
($\varepsilon=0.5$, $M_{\rm BH}=10^8 M_{\odot}$, $f_{\rm X}=0.1$, 
and $\rho_0=10^{-9}\,{\rm g\,cm^{-3}}$).
The black dashed line with the open circles
and the black dashed-doted line with the filled circles
are the same as the red line, but for $\rho_0=10^{-7}\,{\rm g\, cm^{-3}}$ and $\rho_0=10^{-11}\,{\rm g\, cm^{-3}}$.
}\label{Fig7}
\end{figure}

The force multiplier, $M(\xi,t)$, as a function of 
the ionization parameter, $\xi$, and the local optical depth parameter, $t$,
is shown in Figure \ref{Fig7}.
The solid red line with red filled circles 
shows the location in $t$-$\xi$ space 
obtained by
tracing the $R=30R_{\rm S}$ line from 
$z\sim 0$ to $z\sim 0.2$ ($\theta\sim 89.98^{\circ}$)
based on our results.
The red circles roughly move from upper left to lower right
with an increase of the altitude,
since the density is decreasing with $z$,
and since $t$ and $\xi$ are almost proportional to 
$\rho$ and $\rho^{-1}$ (see the equations [\ref{xi}] and [\ref{t-xi}]).
Note that a surge of $t$ at the upper-left region 
is caused by that $|dv_r/dr|$ is very close to null at the disk surface.
The red line passes through the region of $M\gtrsim 0.1$
when $\xi \sim 10$--$100$ and $t \sim 10^{-3}$--$10^{-2}$,
at which enhanced line force assists the matter to move upward. 
%

Near the disk surface ($z\lesssim 0.01 R_{\rm S}$), 
the local optical depth parameter is too large 
for the force multiplier to exceed 0.1.
In this region, the matter is not accelerated by the line force.
The density distribution is adjusted 
so as to keep the balance between the gravitational force and 
sum of the gas pressure force and the radiation force
via the electron scattering.
%
Although the matter moves upward because of $M\gtrsim 0.1$ 
at around $z \sim 0.01$--$0.07R_{\rm S}$,
%
the overionization reduces the force multiplier 
to less than 0.1 at a higher region ($z\gtrsim 0.07R_{\rm S}$). 
Thus, a 'traffic jam' is induced 
by that 
the upward velocity at $z\sim 0.1$--$0.2R_{\rm S}$
decreases.
The density increases by this traffic jam 
so that the ionization parameter 
becomes smaller, making the force multiplier larger again
(see $z\sim 0.2R_{\rm S}$ where
$\log \xi \sim 2.3$, $\log t \sim -3$, and $M \sim 0.6$).
The enhanced line force blows away the high-density matter outward,
which is the origin of the mainstream of the wind.
At $z\gtrsim 0.2R_{\rm S}$,
the matter gradually falls down.

\begin{figure}
  \begin{center}
    \FigureFile(77mm,77mm){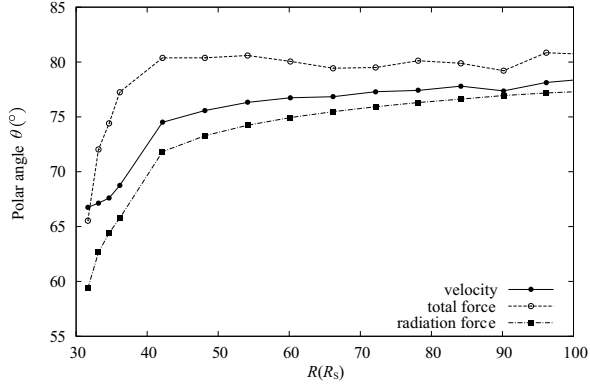}
  \end{center}
  \caption{
Polar angles of the directions of the velocity (solid line with the filled circle), the total force (dashed line with the open circle), and the radiation force (dashed-doted line with the filled square) along the dense streaky structure (see Figure \ref{Fig1})
The total force includes the radiation force, the gravitational force, and the centrifugal force (other than the gas pressure force). 
The horizontal axis is the distance from the central black hole along the disk surface.
}.\label{Fig3}
\end{figure}

Next, we focus on the bending of the mainstream.
In Figure \ref{Fig3}, the directions of the velocity 
(solid line with the filled circles), 
the total force (dashed line with the open circles), 
and the radiation force (dashed-doted line with the filled squares) 
along the mainstream are shown
as a function of 
$R$.
Here, the total force means the sum of the radiation force, 
the gravitational force, and the centrifugal force.
The gas pressure force is negligible.
It is found that the direction of the velocity gradually shifts 
from $\theta\sim 65^{\circ}$ to $\sim 78^{\circ}$.
This means that the matter is lifted relatively upward 
near the wind base 
and the outflow turns in the radial direction with an increase of $R$.
Thus, the mainstream of the wind becomes funnel-shaped.

Such a bending of the stream is caused 
by the radiation force 
in cooperation with the centrifugal force.
Since the radiation comes form the disk surface,
the $\theta$-component of the radiation force
is non-trivial at the inner region,
in contrast, gets close to null at the outer region.
Thus, the direction of the radiation force including the line force
changes from $\theta \sim 60^\circ$ to $\sim 78^\circ$
(almost $r$-direction).
The direction of the total force is different from that of the radiation force at $R\lesssim 50R_{\rm S}$,
but both get closer at $R\gtrsim 50R_{\rm S}$.
This is because the centrifugal force is considerable only near the launching region.
%
Hence, the wind launched towards the direction of $\theta\sim 65^\circ$
at the wind base
gradually changes the direction and finally just flows 
along the direction of the radiation force (total force), 
$\sim 78^{\circ}$.


\section{Dependence on density at disk surface}
\label{App2}
\begin{figure}
  \begin{center}
    \FigureFile(77mm,77mm){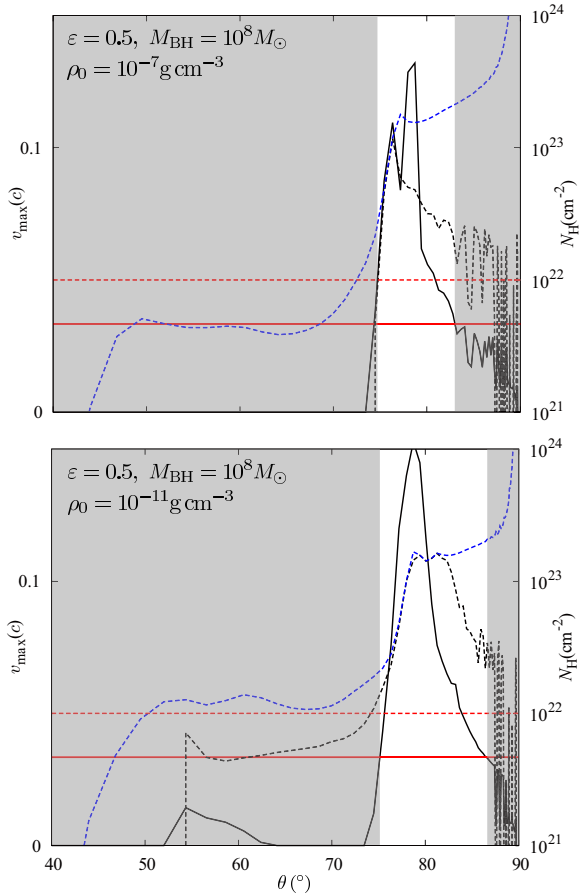}
  \end{center}
  \caption{
Same as the top panel of Figure \ref{Fig2}, but for 
$\rho_0=10^{-7}\,{\rm g\, cm^{-3}}$ (top panel) and for
$\rho_0=10^{-11}\,{\rm g\, cm^{-3}}$ (bottom panel). 
}
\label{Fig6}
\end{figure}

The wind structure does not change so much
if we change $\rho_0$.
%
Figure \ref{Fig6} is the same as Figure \ref{Fig2} 
but for $\rho_0=10^{-7}\,{\rm g\, cm^{-3}}$  (top panel)
and for $\rho_0=10^{-11}\,{\rm g\, cm^{-3}}$  (bottom panel).
In both cases,
the funnel-shaped disk wind with opening angle of 
$\theta \sim 78^\circ$ appears,
and 
the UFO probability is 20\% for $\rho_0=10^{-11}\,{\rm g\, cm^{-3}}$
and 14\% for $\rho_0=10^{-7}\,{\rm g\, cm^{-3}}$.
%
Although the unshaded region is slightly narrow
for $\rho_0=10^{-7}\,{\rm g\, cm^{-3}}$,
we find that the column density and the velocity 
with $2.5 \leq \log \xi < 5.5$ 
are insensitive to $\rho_0$
in the unshaded region
(see Figures \ref{Fig2} and \ref{Fig6}).
%

%
In Figure \ref{Fig7}, the black dashed line with the open circles and 
the black dashed-doted line with the filled circles
show the location in $t$-$\xi$ space obtained by
our calculations
for $\rho_0=10^{-7}\,{\rm g\, cm^{-3}}$
and for $\rho_0=10^{-11}\,{\rm g\, cm^{-3}}$.
%
We find that the force multiplier differs depending on $\rho_0$ 
at the very vicinity of the disk surface.
However, as the density decreases with the altitude,
three lines get close and 
converge 
at $z\gtrsim 0.01R_{\rm S}$. 
This simply explains why
the wind structure as well as the UFO probability 
does not change so much if we change $\rho_0$ in the present simulations.

\end{document}